\documentclass{article}
\usepackage{ijcai24}

\usepackage{times}
\usepackage{soul}
\usepackage{url}
\usepackage[hidelinks]{hyperref}
\usepackage[utf8]{inputenc}
\usepackage[small]{caption}
\usepackage{graphicx}
\usepackage{amsmath}
\usepackage{amsthm}
\usepackage{booktabs}
\usepackage[switch]{lineno}

\usepackage{bm}
\usepackage{subcaption}
\usepackage{amssymb,amsmath,stmaryrd}
\usepackage{booktabs}
\usepackage{xcolor}
\usepackage{threeparttable}
\newtheorem{theorem}{Theorem}
\usepackage{amsfonts}
\usepackage{mathtools}
\usepackage[ruled,linesnumbered]{algorithm2e}
\newcommand*{\defeq}{\stackrel{\text{def}}{=}}
\let\oldnl\nl
\newcommand{\nonl}{\renewcommand{\nl}{\let\nl\oldnl}}
\SetKwInput{KwData}{Input}
\SetKwInput{KwResult}{Output}
\DontPrintSemicolon

\usepackage{tikz}
\DeclareMathOperator*{\bplus}{\tikz[baseline,line width=.1ex]{\node[minimum size=1.1em,draw,anchor=base,inner sep=.1ex]{$+$};}}

\newcommand{\subfigwidth}{0.246}

\newcommand*{\rand}{\stackrel{\text{\$}}{\longleftarrow}}

\title{Skefl: Single-Key Homomorphic Encryption for Secure Federated Learning}

\author{
    Dongfang Zhao
    \affiliations
    University of Washington
    \emails
    dzhao@uw.edu
}

\begin{document}

\maketitle

\begin{abstract}
Homomorphic encryption (HE) is widely adopted in untrusted environments such as federated learning. A notable limitation of conventional single-key HE schemes is the stringent security assumption regarding collusion between the parameter server and participating clients: Adversary clients are assumed not to collude with the server, as otherwise, the parameter could transmit the ciphertext of one client \(C_0\) to another client \(C_1\), who shares the same private key and could recover the local model of \(C_0\). One plausible solution to alleviate this strong assumption is multi-key HE schemes, which, unfortunately, prove impractically slow in production systems.
In this work, we propose a new protocol that achieves the balance between security and performance: We extend single-key HE schemes with efficient secret sharing, ensuring that collusion between the parameter server and any compromised clients cannot reveal any local model. We term this protocol Skefl: Single-key homomorphic encryption for secure federated learning. The key idea behind Skefl is the secret-sharing of homomorphic \textit{ciphertexts} generated by multiple clients using the same pair of secret and public keys. We will substantiate the security claims of the proposed protocol using the well-known simulation framework in cryptography. Additionally, we will report on the practical performance of the Skefl protocol.
\end{abstract}

\section{Introduction}

\paragraph{Background}

The application of homomorphic encryption (HE)~\cite{cgentry_stoc09} to federated learning (FL)~\cite{mcmahan_aistat17} has garnered significant research attention. HE enables the parameter server to aggregate encrypted local models without gaining knowledge of model weights, thereby preserving the privacy of sensitive training data~\cite{lzhu_nips19}.
Among existing HE schemes~\cite{bfv,bgv,ckks}, CKKS~\cite{ckks} is considered the most promising, supporting approximate encryption on floating-point numbers. Furthermore, CKKS facilitates the encryption of vectors, a highly desirable feature for machine learning workloads. Notable libraries for CKKS and other HE schemes include HElib~\cite{helib}, SEAL~\cite{sealcrypto}, and OpenFHE~\cite{openfhe}.


\paragraph{Motivation}
The utilization of the conventional Homomorphic Encryption (HE) scheme in Federated Learning (FL) presents challenges in both security and performance.

\begin{itemize}
    \item \textbf{Security:} The conventional HE scheme issues a pair of private and public/evaluation keys, enabling untrusted parties to use the public/evaluation key for algebraic operations on ciphertexts, while data owners use the private key to recover the algebraic results. In FL, however, this necessitates that all clients share the same private key. If the parameter server (PS) and any semi-honest client collude, the local model's weights can be compromised: the PS provides the ciphertext, and the semi-honest client provides the private key.

    \item \textbf{Performance:} One plausible solution to the aforementioned security issue is to adopt multi-key HE schemes~\cite{alopez_stoc12,hchen_ccs19,tkim_ccs23}. The concept of multi-key HE (MKHE) involves every client generating a distinct pair of private and public keys, such that the algebraic operations on ciphertexts incorporate $n$ public/evaluation keys, where $n$ denotes the total number of parties. Consequently, the ciphertext result can only be decrypted with all the $n$ private keys. Thus, collusion between the parameter server and a subset of compromised clients (i.e., not all the $n$ clients are compromised, in which case the system cannot be secure anyway) cannot recover the original model weights. However, the most well-known MKHE scheme~\cite{tkim_ccs23} incurs computational costs with a linear growth in the number of clients, exacerbating the performance challenge of (single-key) HE~\cite{otawose_sigmod23} and rendering it impractical for federated learning, where the number of clients could reach tens of millions on mobile devices~\cite{mcmahan_aistat17}.
\end{itemize}

\subsection{Proposed Approach}




In this work, we introduce a novel protocol designed to achieve a balance between security and performance. We extend single-key homomorphic encryption (HE) schemes with efficient secret sharing to prevent collusion between the parameter server and compromised clients from revealing any local model. We refer to this innovative protocol as Skefl: Single-key homomorphic encryption for secure federated learning. The core concept of Skefl lies in the secret-sharing of homomorphic \textit{ciphertexts} generated by multiple clients using the same pair of secret and public keys. We term this secret-sharing of ciphertext as \textit{asymmetric threshold secret sharing} (ATSS), signifying that the parameter server and compromised clients cannot recover the local model without the ciphertext share of the client who owns it.

We emphasize key distinctions between Skefl and existing protocols for applying HE schemes in FL systems:

\begin{itemize}
    \item Skefl employs secret sharing of ciphertext messages, where the homomorphic aggregation of ciphertext shares equals the original encryption of a local model. This is distinct from the secret sharing of plaintext models, where reconstruction can be trivially achieved through arithmetic summation.

    \item Skefl necessitates homomorphic encryption of (shares of) local models, as the parameter server could otherwise unilaterally recover the global model. That is, a straightforward secret-sharing scheme of plaintext models is insufficient.
\end{itemize}

\subsection{Contributions}

In summary, this paper's technical contributions include:
\begin{itemize}
    \item We present a new protocol, Skefl, designed for the efficient and secure application of homomorphic encryption in federated learning systems.
    \item We establish the provable security of Skefl using the classical simulation framework in cryptography.
    \item We implement Skefl within a practical federated learning framework tailored for collaborative scientific computing, presenting promising results across various benchmarks.
\end{itemize}


\section{Preliminaries and Related Work}

\subsection{Secure Aggregation and Poisoning Attacks}

Secure aggregation algorithms often assume a centralized parameter server for effective collaboration. In a study by~\cite{dyin_icml18}, robust distributed gradient descent algorithms rooted in median and trimmed mean operations were analyzed. Additionally,~\cite{lli_aaai19} introduced a class of robust stochastic sub-gradient methods for distributed learning from heterogeneous datasets while addressing an unknown number of adversarial workers, with a master machine assumption. These algorithms typically operate under the \textit{semi-honest} (or \textit{honest-but-curious}) attack model, wherein participants adhere to the protocol but may passively probe, store, or analyze data. The semi-honest model is a prevalent threat assumption in production systems~\cite{zhang_atc20}.

Beyond the semi-honest model, another crucial perspective assumes \textit{malicious} participants who could deviate from the protocol and even manipulate data. For instance, the deliberate introduction of compromised samples to the training dataset constitutes a \textit{data poisoning} attack~\cite{alfeld2016}. Similarly, a malicious client might upload an arbitrary model, referred to as a \textit{model poisoning} attack~\cite{li2020review}. Numerous well-known attacks fall within this category, including backdoor attacks~\cite{yang2019,bhagoji2019analyzing,BagdasaryanVHES20,Xie2020DBA}. The security community has proposed diverse solutions to counter these poisoning attacks. In~\cite{mfang_security20}, methods to poison clients' local models and corresponding defense strategies were discussed. Recent research along this trajectory includes~\cite{ddata_icml21,yyang_icml21,sun2021,Shejwalkar2021}, further advancing the understanding and mitigation of these security threats.

\subsection{Homomorphic Encryption and Secret Sharing}

An encryption function $f()$ is deemed \textit{additive homomorphic} when the equation $\displaystyle f^{-1}\left(f(a) \oplus f(b)\right) = a + b$ holds, where $f^{-1}()$ denotes decryption and $\oplus$ signifies the binary operation within the range of $f()$. A parallel property exists for the multiplicative operation. When a function supports both additive and multiplicative homomorphism, it is termed \textit{fully homomorphic encryption} (FHE)~\cite{cgentry_stoc09}. While diverse FHE schemes like BFV~\cite{bfv} and CKKS~\cite{ckks} have been proposed, their broader adoption in HEFL is hindered by substantial computational overhead. Instead, HE schemes supporting only additive (or multiplicative) homomorphism, such as Paillier encryption~\cite{ppail_eurocrypt99}, have found efficiency in PPML systems~\cite{symmetria_vldb20}, including adoption in works like~\cite{shardy_arxiv17,zhang_atc20}.

Conversely, information-theoretic techniques like threshold secret sharing (TSS)~\cite{ashamir_cacm79,trabin_stoc89} offer an alternative to cryptographic methods. TSS protects data confidentiality by fragmenting a message into $n$ trunks, each assigned to one of $n$ participants. A $(t,n)$ TSS protocol ensures that $t$ or more chunks suffice to reconstruct the original message, while $t$ or fewer chunks reveal no original message information. Despite various implementations of $(t,s)$ TSS protocols, like the Lagrange interpolating polynomial and the Chinese Remainder Theorem (CRT), the core concept remains consistent—sufficient participants collectively and uniquely solve a system of equations. TSS serves as a fundamental element in secure multiparty computation (MPC), facilitating collaborative computation of functions by a group of participants, such as aggregating local models in PPML and FL. MPC finds application in multiple PPML frameworks like DeepSecure~\cite{deepsecure}, SecureML~\cite{secureml}, and ABY~\cite{aby}.

\section{Methodology}
\subsection{ATSS: Asymmetric Threshold Secret Sharing}

\subsubsection{Notations}
Let $f > 0$ denote the maximal number of semi-honest clients in an FL system,
in which there are a single semi-honest server $\mathcal{S}$ and a set of $n$ clients $\mathcal{C} = \{C_1, C_2, \dots, C_n\}$.
Therefore, there are at least $(n-f)$ honest clients.
We assume that the majority of clients are honest,
implying that $2f < n$.
If the context is clear, we will use $2f + 1 = n$ to replace the above inequality;
this is just to simplify the analysis a bit and is not a technical requirement.
This also implies that $\mathcal{S}$ can collude with up to $f$ clients.
We use $\textbf{W}_i$ to denote the local model (weights) trained at client $C_i$.
In production systems, a model usually comprises layers of weights and biases;
in our discussion, we simply the application-specific implementation by a real-valued vector $\textbf{W}$.
Since the primitive somehow decompose $\textbf{W}_i$ into a list of sub-models,
we use superscripts to denote the sub-models:
$\textbf{W}_i^j$ denotes the $j$-th sub-model, or trunk, 
of $\textbf{W}_i$, $1 \le j \le i$.
Obviously, $\textbf{W}_i = \textbf{Rec}_{j=1}^i (\textbf{W}_i^j)$,
where $\textbf{Rec}()$ denotes the reconstruction function over sub-models to reconstruct the original model.
Without loss of generality, we can think of $C_j$ stores the $j$-th sub-model of the original model trained by $C_i$.
By convention, we use $d \rand \bm{U}$ to denote that an element $d$ is randomly selected from set $\bm{U}$.
We use $\mathbb{Z}^+$ to denote the set of positive integers.

\subsubsection{Decomposition}
Each $\textbf{W}_i$ is decomposed into $(f+1)$ sub-models.
We choose $(f+1)$ to reflect the balance between security and performance.
If the number of sub-models is less than $(f+1)$,
it is then possible for $f$ clients to jointly recover the ciphertext of $\textbf{W}_j$.
On the other hand, the overall number of chunks being transmitted in the entire FL system,
and thus the network overhead thereof,
is proportional to the number of sub-models;
therefore, we want to keep the number of sub-models as small as possible.
As a result, $(f+1)$ is the minimal number that guarantees the confidentiality of the primitive.

The indices of clients (i.e., $j$'s) to receive the sub-models from $\textbf{W}_i$ are chosen randomly except for $C_i$ itself.
That is, $C_i$ will always hold a chunk and there are additional $f$ clients holding the remaining $f$ chunks. 
As a result, there are ${n \choose f}$ possibilities for the set of clients assigned to hold the chunks of a specific local model.
In contrast with the conventional threshold secret-sharing (TSS) scheme,
client $C_i$ keeps not only $\textbf{W}_i^k$, 
$k \in J$ where $J$ denotes the set of client indices $j$'s,
but also the intact $\textbf{W}_i$.
Therefore, $C_i$ has an asymmetric role in the above secret-sharing scheme:
$C_i$ could re-decompose $\textbf{W}_i$ by canceling out the previous trunks and re-broadcasting the newly split chunks to a different set of $j$'s.
We call the above primitive \textit{asymmetric threshold secret-sharing} (ATSS).
The point of keeping the original $\textbf{W}_i$ will be further elaborated on later in the security analysis.

ATSS decomposes $\textbf{W}_i$ using an additive secret-sharing scheme into a set of shares stored locally on the current client.
Unsurprisingly, the share assigned to $C_i$ itself is calculated as follows:
\[\displaystyle
enc(w^i) = \left(\bigoplus_{j \not= i} enc(w)^j\right) \oplus enc(w),
\]
where $enc$ denotes the homomorphic encryption, $\oplus$ denotes the homomorphic addition operation between two encrypted models, and $\bigoplus$ denotes the consecutive $\oplus$'s among a set of ciphertext models.

The complexity of ATSS decomposition is as follows.
Let $m$ denote the cardinality of tensor $W_i$ in the FL model.
Each client sends out $m \cdot f$ messages to the system.
Therefore, in each FL training round, there are a total of $\mathcal{O}(nmf)$ messages.
If we assume $2f + 1 = n$,
then the overall message complexity of each FL round is $\mathcal{O}(mn^2)$.
The overall computational complexity is also $\mathcal{O}(mn^2)$:
there are $n$ clients,
each of which computes the $\oplus$ operation $f$ times (i.e., $\mathcal{O}(n)$) for each of $m$ elements.

\subsubsection{Reconstruction}

The reconstruction from the chunks is straightforward because $\oplus$ is symmetric.
To reconstruct $\textbf{W}_i$, ATSS computes each of its elements $w$ as follows:
\[
enc(w) = \bigoplus_{j \in J} enc(w)^j,
\]
where $j$ denotes the $j$-th chunk of $\textbf{W}_i$ and $J$ is the set of chunk indices.
The requester can convert the bit string back to a float number as an element of the original vector.
Consequently, the aggregation function $\textbf{Rec}()$ can be defined as the extension of the above element-wise operation.

The complexity of ATSS aggregation is similar to decomposition.
Both the time and message complexity are $\mathcal{O}(mn^2)$ because all the operations are exactly the reverse of decomposition.

\subsubsection{Verification}

For clients or server $\mathcal{S}$ who wants to verify the validity of an encrypted $enc(W_i)$ generated by client $C_i$,
$C_i$ publishes the hash value of $W_i$, $H(enc(W_i))$,
where $H()$ denotes a cryptographic hash function, such as SHA256.
Please note that the verification is for the completeness of the ATSS primitive and unnecessarily involved in the FL workloads.

The message complexity of verifying a single model is $\mathcal{O}(n)$
because the requester would need to collect $f$ chunks from other clients with $f$ messages.
Since we assume $2f + 1 = n$, 
it implies that the message complexity is linear to the number of clients.
The time complexity of verifying a single model, however,
is constant $\mathcal{O}(1)$ because the requester only needs to compare the published hash value of the model with the locally reconstructed one.

\subsubsection{Primitives}

We conclude this section with a list of formally defined ATSS primitives as follows.
\begin{itemize}
    \item \textbf{ATSS.Split}() is a function:
    \[\displaystyle
        \{0, 1\}^\ell \to \left( \{0, 1\}^\ell \right)^{f+1},
    \] 
    where the input \textit{in} is a string of length $l$ (in bits) and the output \textit{out} is $(f+1)$ $l$-bit strings whose XOR equals the input.
    
    \item \textbf{ATSS.Merge}() is a function:
    \[\displaystyle
        \left( \{0, 1\}^\ell \right)^{f+1} \to \{0, 1\}^\ell ,
    \] 
    where the input \textit{in} is a set of $l$-bit strings and the output \textit{out} is a single $l$-bit string.
    Note that \textbf{ATSS.Split}(\textit{x}) = \textit{y} $\implies$ \textbf{ATSS.Merge}(\textit{y}) = \textit{x},
    and the converse does not hold in general.
    
    \item \textbf{ATSS.Verify}() is a function:
    \[\displaystyle
        \{0, 1\}^\ell \times \mathbb{Z}^+ \to \{0, 1\},
    \] 
    where the first input is the encryption of a model $enc(W_i)$,
    the second input is the client index verifying the encrypted model,
    and the output indicates success (1) or failure (0).

\end{itemize}

\subsection{Secure Aggregation with Homomorphic Secret-Sharing Ciphertext}

This section depicts the protocol adopting the ATSS primitive into a single-key homomorphic encryption scheme for federated learning, namely Skefl.
A number of building blocks of Skefl are built upon the ATSS primitives.
We start by the extended primitive of \textbf{ATSS.Split}().

\subsubsection{Weighted Sub-models for Non-IID Data}
While the vanilla \textbf{ATSS.Split}() can decompose a single vector,
it has to be extended to take care of the data heterogeneity in federated learning,
also known as not independent and identically distributed (non-IID).
For instance, in the original aggregation algorithm called \textbf{FedAvg}(), the global model is calculated as the weighted average of all local models:
\[\displaystyle
\bm{W} = \textbf{FedAvg}(1, n, \bm{W}_i) \defeq \sum_{i=1}^{n} \frac{N_i}{N} \bm{W}_i,
\]
where $N$ denotes the total number of data samples, 
$N_i$ denotes the number of data samples on client $C_i$,
and $n$ denotes the number of clients.

The extended decomposition algorithm is presented in Alg.~\ref{alg:fedsh_split}.
Line 2 calls the \textbf{ATSS.Split}() primitive to generate a list of $f$ random sub-models and one calculated sub-model.
It does not matter which one of the $(f+1)$ is the calculated one because the remainder of the algorithm will assume the first $f$ sub-models will be adjusted and sent to $f$ distinct clients than $C_i$---the host where the algorithm is being executed.
Line 5 modifies each of the $f$ sub-models by calling the aggregation function on the sub-model itself.
This can be trivially done by setting the first two arguments as identical integers,
such as $i$s.
The effect of doing so is to adjust the weight of each sub-model;
without this step, the global aggregation with Skefl would be inconsistent with the vanilla \textbf{Aggr}() function.
Lines 6--8 pick an unused client $C_d$ and send the modified sub-model to it.
Finally, client $C_i$ calculates its own sub-model that will be sent to the aggregator.

The complexity of Alg.~\ref{alg:fedsh_split} is as follows.
Lines 4--9 take $\mathcal{O}(n)$ iterations,
each of which incurs the aggregation that costs another $\mathcal{O}(n)$.
Line 2 itself costs $\mathcal{O}(mn^2)$.
Therefore, the overall time complexity of Alg.~\ref{alg:fedsh_split} is $\mathcal{O}(mn^3)$ when \textbf{ATSS.Split}() is used as a black-box function.
However, if \textbf{ATSS.Split}() is open-source then the key modifications in Lines 5--7 can be integrated into the internals of \textbf{ATSS.Split}(),
which would make Alg.~\ref{alg:fedsh_split} result in the same asymptotic complexity of \textbf{ATSS.Split}(),
i.e., $\mathcal{O}(mn^2)$.

\begin{algorithm}
\SetAlgorithmName{Algorithm}{}{}
\caption{Skefl Distribute \textbf{Skefl\_Dist}()}\label{alg:fedsh_split}
\KwData{
A local model $\bm{W}_i$ trained by $C_i$;
The aggregation algorithm \textbf{Aggr}(1, $n$, $\bm{W}_i$) taken by the FL system (e.g., \textbf{FedAvg});
The number of semi-honest clients $f$;
The total number of clients $n$;
The bit-string length of data item $l$;
}
\KwResult{
A set of sub-models $enc(\bm{W}_i^j)$, $1 \le j \le f+1$, satisfying \textbf{Aggr}(enc($\bm{W}_i$)) = $enc(\bm{W})$; 
}
\For{$i = 1; i \le n; i++$}{
    $(enc(\bm{W}_i^1), \dots, enc(\bm{W}_i^{f+1})) =$ \textbf{ATSS.Split}($\bm{W}_i$)\;
    $\bm{U} = \{1, \dots, n\} \setminus \{i\}$ \;
    \For{$j = 1; j \le f; j++$}{
        $enc(\bm{W}_i^j) = \textbf{Aggr}(j,j,enc(\textbf{W}_i^j))$ \;
        $d \rand \bm{U}$ \;
        $C_i$ sends $enc(\bm{W}_i^j)$ to $C_d$ \;
        $\bm{U} = \bm{U} \setminus \{d\}$ \;
    }
    $\displaystyle enc(\bm{W}_i^{f+1}) = \left( \bigoplus_{j=1}^f enc(\bm{W}_i^j) \right) \oplus \textbf{Aggr}(1,1,enc(\bm{W}_i))$\;
}

\end{algorithm}
    
\subsubsection{Garbled Homomorphism of Local Models}

After each client $C_i$ runs \textbf{Skefl\_Dist()},
a local aggregation is performed to strengthen the confidentiality of the original model $\bm{W}_i$.
The goal of doing so is to garble the sub-models generated by \textbf{Skefl\_Dist}() such that semi-honest clients cannot collude with the semi-honest server to possibly identify the sub-models. 
Specifically, each client $C_i$ applies the aggregation function \textbf{Aggr}() over all of the sub-models it receives (including the one sent by itself).
Formally, we defined the garbled local model $\widehat{\bm{W}}_i$ as
\[
\widehat{\bm{W}}_i \defeq \textbf{Aggr}(1, n, \bm{W}_k^*),
\]
where we use $^*$ to denote an arbitrary index of sub-model from $C_k$.
If there are no sub-models sent from $C_k$ to $C_i$, 
we set the default value as the zero vector $\bm{W}_k^* = \bm{0}$.

Client $C_i$ then applies the homomorphic encryption algorithm $\textbf{Enc}$() to $\widehat{\bm{W}}_i$ to even protect the garbled (summation of) sub-models.
We define a function for the local garbling aggregation as a function:
\begin{equation}\label{eq:fsh_garble}
\displaystyle
\textbf{Skefl\_Garble}(i) \defeq \textbf{enc}(\widehat{\bm{W}}_i).
\end{equation}
On average, each client $C_i$ would receive $n$ sub-models and the local aggregation involves up to $nm$ operations.
Therefore, the time complexity of \textbf{Skefl\_Garble}() is $\mathcal{O}(mn)$.

\subsubsection{Secure Aggregation with Skefl}

The server $\mathcal{S}$ receives from each client $C_i$ the encryption of a garbled model $\textbf{enc} (\widehat{\bm{W}}_i)$.
Because \textbf{enc}() is homomorphic,
the aggregation function \textbf{Aggr}() can be converted into a series of algebraic operations over the ciphertexts.
Recall that a homomorphic function implies that:
\[
\textbf{dec}_{sk}(\textbf{enc}_{pk}(x) \boxplus \textbf{enc}_{pk}(y)) = x + y,
\]
where \textbf{dec}() is the reverse function of homomorphism \textbf{enc}().
Therefore, if we apply the corresponding aggregation over the ciphertexts,
the result would be decrypted into the aggregation of the plaintext,
which is exactly what the clients compute with the private/secret key $sk$.
If we extend the homomorphic addition $\boxplus$ to a set of inputs,
then $\displaystyle \bplus_{i = 1}^n \textbf{enc}_{pk}(\widehat{\bm{W}}_i)$ denotes the consecutive summation of the set of cryptographically garbled models from all clients.

Let \textbf{Skefl\_Aggr}() denote the above equation:
\[\displaystyle
\textbf{Skefl\_Aggr}(\{\bm{W}_i\}) \defeq \bplus_{i = 1}^n \textbf{enc}_{pk}(\widehat{\bm{W}}_i),
\]
where $1 \le i \le n$.
We claim that 
\begin{equation}\label{eq:fsh_aggr}
\displaystyle
\textbf{dec}_{sk}(\textbf{Skefl\_Aggr}(\{\bm{W}_i\})) = \textbf{Aggr}(1, n, \bm{W}_i).    
\end{equation}
We skip the formal proof and only highlight the key step for verifying the above equation (omitting the $pk$ and $sk$ subscripts):
\[\displaystyle
\textbf{dec}\left(\bplus_{i = 1}^n \textbf{enc}\left(\widehat{\bm{W}}_i\right)\right) 
=
\textbf{dec}\left(\textbf{enc}\left(\textbf{Aggr}(1,n,\bm{W}_i)\right)\right),
\]
which essentially states that we can push the aggregation inside the ciphertext due to the homomorphic property.

The time complexity of \textbf{Skefl\_Aggr}() is straightforward.
The naive way of applying $\bplus$ is obviously $\mathcal{O}(n)$.
However, if we can leverage multiple cores to parallelize the summation through a binary tree,
then the time complexity can be reduced to $\mathcal{O}(\log n)$ in practical systems.

\subsection{Security Analysis}

\subsubsection{Threat Model}

We assume all adversaries are \textit{efficient},
meaning that they run polynomial algorithms in the bit length of the input.
Roughly speaking, this implies that the adversaries do not have access to unlimited computational power.
We also assume all adversaries are \textit{semi-honest},
which is also known as \textit{honest-but-curious}.
This entails two assumptions:
\begin{itemize}
    \item All clients $C_i$'s and the server $\mathcal{S}$ follow the aggregation algorithm \textbf{Aggr}() without tempering with the models $\bm{W}_i$'s;
    \item Some (up to $f$) clients and the server $\mathcal{S}$ may probe and store data during the execution of the FL workloads.
\end{itemize}
Specifically, server $\mathcal{S}$ could collude with a client $C_j$ to hopefully recover a local model trained by another client $C_i$, $j \not= i$.
We will show that Skefl can resist such attacks in the next section.

We assume the building block of the homomorphic encryption scheme is \textit{computational} secure under the \textit{chosen-plaintext attack} (CPA).
In practice, there are multiple CPA-secure schemes,
such as Paillier~\cite{ppail_eurocrypt99} and Symmetria~\cite{symmetria_vldb20}.
There are two ways to define the \textit{indistinguishability} under CPA (IND-CPA) when an adversary $\mathcal{A}$ obtains a polynomial number of plaintext-ciphertext pairs $(m, \textbf{enc}(m))$'s:
(i) $\mathcal{A}$ cannot distinguish a specific ciphertext $\textbf{enc}(m)$ from a random string; and
(ii) $\mathcal{A}$ cannot distinguish $\textbf{enc}(m_1)$ from $\textbf{enc}(m_2)$, $m_1 \not= m_2$.
The former definition implies the latter one,
although the latter definition is more widely adopted.
In this work, we will take the second definition.
That is, we assume the probability for $\mathcal{A}$ to guess the right message from $\textbf{enc}(m_1)$ and $\textbf{enc}(m_2)$ is \textit{negligibly} higher than $\frac{1}{2}$ after learning about a polynomial number of $(m, \textbf{enc}(m))$ pairs.
By \textit{negligibly}, we mean the advantage of $\mathcal{A}$ is a function that decreases faster than the inverse of \textit{any} polynomial functions. 

\subsubsection{IND-CPA of Skefl}

Let $l$ denote the bit-string length of the model values.
Without loss of generality, we assume two clients $C_1$ and $C_2$ upload two (encrypted and garbled) models $\bm{W}_1$ and $\bm{W}_2$ to the aggregator.
Let $o$ denote the output by the adversary $\mathcal{A}$,
indicating $\mathcal{A}$'s guess about whether the original plaintext model is from either $C_1$ or $C_2$.
If by assumption \textbf{enc}() is IND-CPA,
then the following holds:
\begin{equation}\label{eq:hecpa}
\displaystyle
\begin{cases}
    Pr\left[o = 1 | \textbf{enc}(\bm{W}_1)\right] \le \frac{1}{2} + \mathcal{O}(\frac{1}{2^l})\\
    Pr\left[o = 2 | \textbf{enc}(\bm{W}_2)\right] \le \frac{1}{2} + \mathcal{O}(\frac{1}{2^l})\\
    Pr[\textbf{enc}(\bm{W}_1)] = Pr[\textbf{enc}(\bm{W}_2)] = \frac{1}{2}
\end{cases}
\end{equation}
where $Pr$[\;] denotes the probability (under the condition of observing a ciphertext).
The last term $\mathcal{O}(\frac{1}{2^l})$ is often called a \textit{negligible function} in the cryptographic literature.

Our goal is to show that the encrypted garbling of $\bm{W}_i$'s are IND-CPA.
That is, the best $\mathcal{A}$ can do is the following:
\begin{equation}\label{eq:fsh_garble_cpa}
\displaystyle
\begin{cases}
Pr\left[o = 1 | \textbf{Skefl\_Garble}(1)\right] \le \frac{1}{2} + \mathcal{O}(\frac{1}{2^l})\\
Pr\left[o = 2 | \textbf{Skefl\_Garble}(2)\right] \le \frac{1}{2} + \mathcal{O}(\frac{1}{2^l})\\
Pr[\textbf{Skefl\_Garble}(1)] = Pr[\textbf{Skefl\_Garble}(2)] = \frac{1}{2}
\end{cases}
\end{equation}
after sending $\bm{W}_1$ and $\bm{W}_2$ to the semi-honest aggregator.
To prove the above equation, 
we will use the popular \textit{reduction} technique,
which is essentially proof by contradiction.

\begin{theorem}[Skefl is IND-CPA]
If \textbf{enc}() is IND-CPA, then \textbf{Skefl\_Garble}() defined in Eq.~\eqref{eq:fsh_garble} is IND-CPA.
\end{theorem}

\begin{proof}[Proof (sketch)]
We use the standard simulation framework to prove the IND-CPA of the proposed Skefl protocol. Suppose there is a perfectly secure oracle that can privately take in a local model for aggregation and would never collude with any clients. 
Without loss of generality, 
we will focus on the $o = 1$ case as $o = 2$ is similar.
Let $\mu$ denote the negligible (asymptotic) function $\mathcal{O}(\frac{1}{2^l})$.
Because 
\[
Pr[o=1 | \textbf{enc}(\bm{W}_1)] \le \frac{1}{2} + \mu,
\]
we know
\[
\frac{Pr[o=1 \wedge \textbf{enc}(\bm{W}_1)]}{Pr[\textbf{enc}(\bm{W}_1)]} \le \frac{1}{2} + \mu,
\]
which is equivalent to
\begin{equation}\label{eq:fsh_proof_cond}
\displaystyle
Pr[o=1 \wedge \textbf{enc}(\bm{W}_1)] \le \frac{1}{4} + \frac{\mu}{2}.   
\end{equation}
Now, suppose 
\[
Pr\left[o = 1 | \textbf{Skefl\_Garble}(1)\right] > \frac{1}{2} + \mu,
\]
which means (similar to before):
\[
Pr\left[o = 1 \wedge \textbf{Skefl\_Garble}(1)\right] > \frac{1}{4} + \frac{\mu}{2}.
\]
According to Eq.~\eqref{eq:fsh_garble}, the following holds:
\[
Pr\left[o = 1 \wedge \textbf{enc}(\widehat{\bm{W}}_1)\right] > \frac{1}{4} + \frac{\mu}{2}.
\]
From the perspective of $\mathcal{A}$ who outputs $o=1$,
the encryption of the original local model $\bm{W}_i$ is indistinguishable from the encryption of the garbled local model $\widehat{\bm{W}}_i$ (required by the IND-CPA property of the encryption).
Therefore, the last equation implies a contradiction to Eq.~\eqref{eq:fsh_proof_cond},
which completes the proof.
\end{proof}

\section{Evaluation}

\subsection{Implementation}

We have implemented the proposed Skefl protocol on top of FedML~\cite{fedml} using the message-passing interface (MPI)~\cite{openmpi} interface.
FedML supports multiple geographically-distributed clusters under the same namespace (i.e., an \textit{MPI\_COMM\_WORLD}) and yet allows each client to be assigned with a single CPU core (i.e., a \textit{rank} in MPI) to save cost.
The early release of FedML does not support a persistence layer for intermediate data;
an experimental feature of blockchain-based data provenance for intermediate models is enabled in our evaluation.
FedML is a pure Python implementation and invokes low-level C++ MPI calls through mpi4py~\cite{mpi4py}.
The homomorphic computation over tensors is implemented through TenSEAL~\cite{tenseal} and we pick the CKKS~\cite{ckks} scheme for our implementation because it is the only one currently supporting homomorphic encryption over float numbers.


\subsection{Experimental Setup}

\subsubsection{Test Bed}
We have deployed Skefl to CloudLab~\cite{cloudlab}. 
Our testbed is a cluster of 10 nodes,
each of which has 40 Intel Xeon Silver 4114 cores at 2.20 GHz, 192 GB ECC DDR4-2666 Memory, and an Intel DC S3500 480 GB SSD.
All nodes are connected to two networks:
a 10-Gbps experiment network (Dual-port Intel X520-DA2 10 Gb NIC) and a 1-Gbps control network (Onboard Intel i350 1 Gb).
We use the experiment network for evaluation.
All experiments are repeated at least three times,
and we report the average numbers along with the standard errors.

\subsubsection{Federated Learning (Hyper-)Parameters}

As for the setting of federated learning,
we set the numbers of local and global epochs both as 10.
The fraction of clients for each round of model updating is 0.1.
The local batch size is set to 50.
The learning rate is 0.05 and the SGD momentum is 0.8.
The number of clients is equal to the number of nodes in our test bed, 10.
The aggregation algorithm is the original FedAvg~\cite{mcmahan_aistat17}.

\subsubsection{Local Machine Learning Models}

We pick two machine learning models to train local neural networks:
convolutional neural network (CNN) and multi-layered perception (MLP).
The CNN model comprises two convolutional layers:
the first being from input data to 10 with kernel size 5 and the second being from 10 to 20 with kernel size 5.
The CNN model then applies two linear layers:
320 $\to$ 50 $\to$ 10.
For the MLP model, there is a 64-neuron hidden layer:
784 $\to$ 64 $\to$ 10.
Both models use ReLU and SoftMax as the activation functions.

\subsubsection{Workloads}

We chose four of the most widely used data sets:
MNIST~\cite{mnist}, Fashion-MNIST~\cite{fmnist}, CIFAR-10~\cite{cifar},
and extended SVHN-extra~\cite{svhn},
all of which can be downloaded from PyTorch.

\subsection{Experimental Results}

\begin{figure*}[!t]
     \centering
     \begin{subfigure}[b]{\subfigwidth\textwidth}
         \centering
         \includegraphics[width=\textwidth]{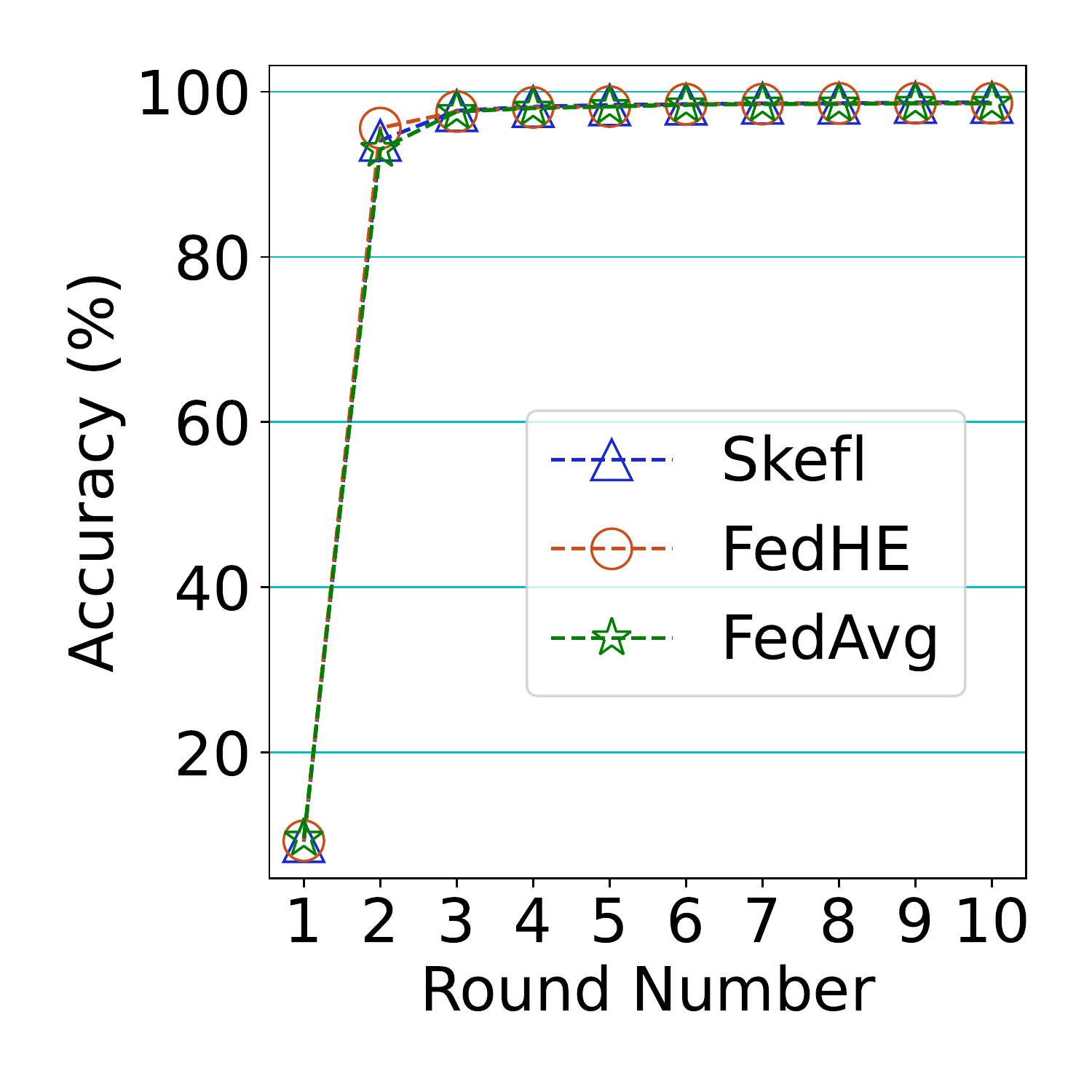}
     \end{subfigure}
     \hfill
     \begin{subfigure}[b]{\subfigwidth\textwidth}
         \centering
         \includegraphics[width=\textwidth]{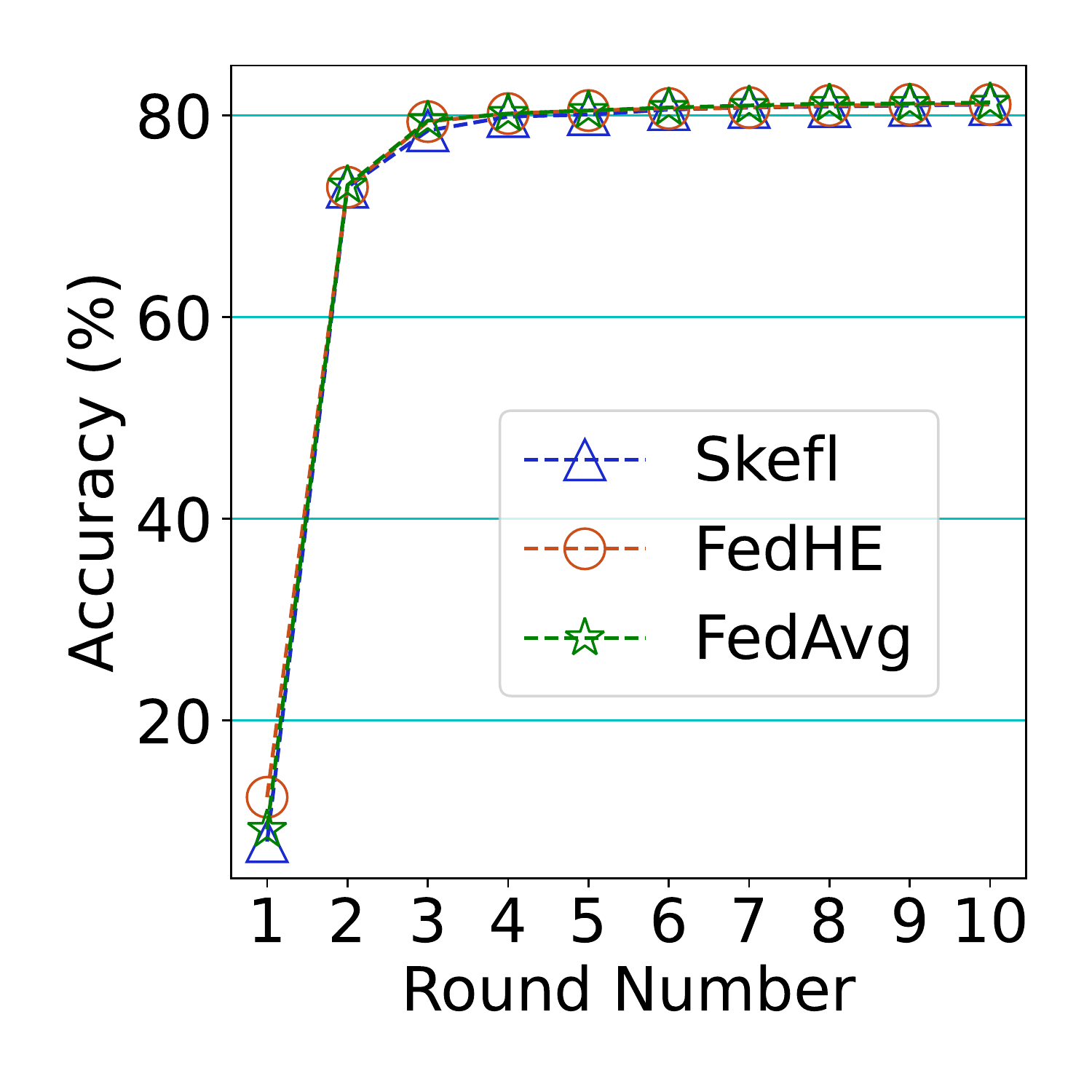}
     \end{subfigure}
     \hfill
     \begin{subfigure}[b]{\subfigwidth\textwidth}
         \centering
         \includegraphics[width=\textwidth]{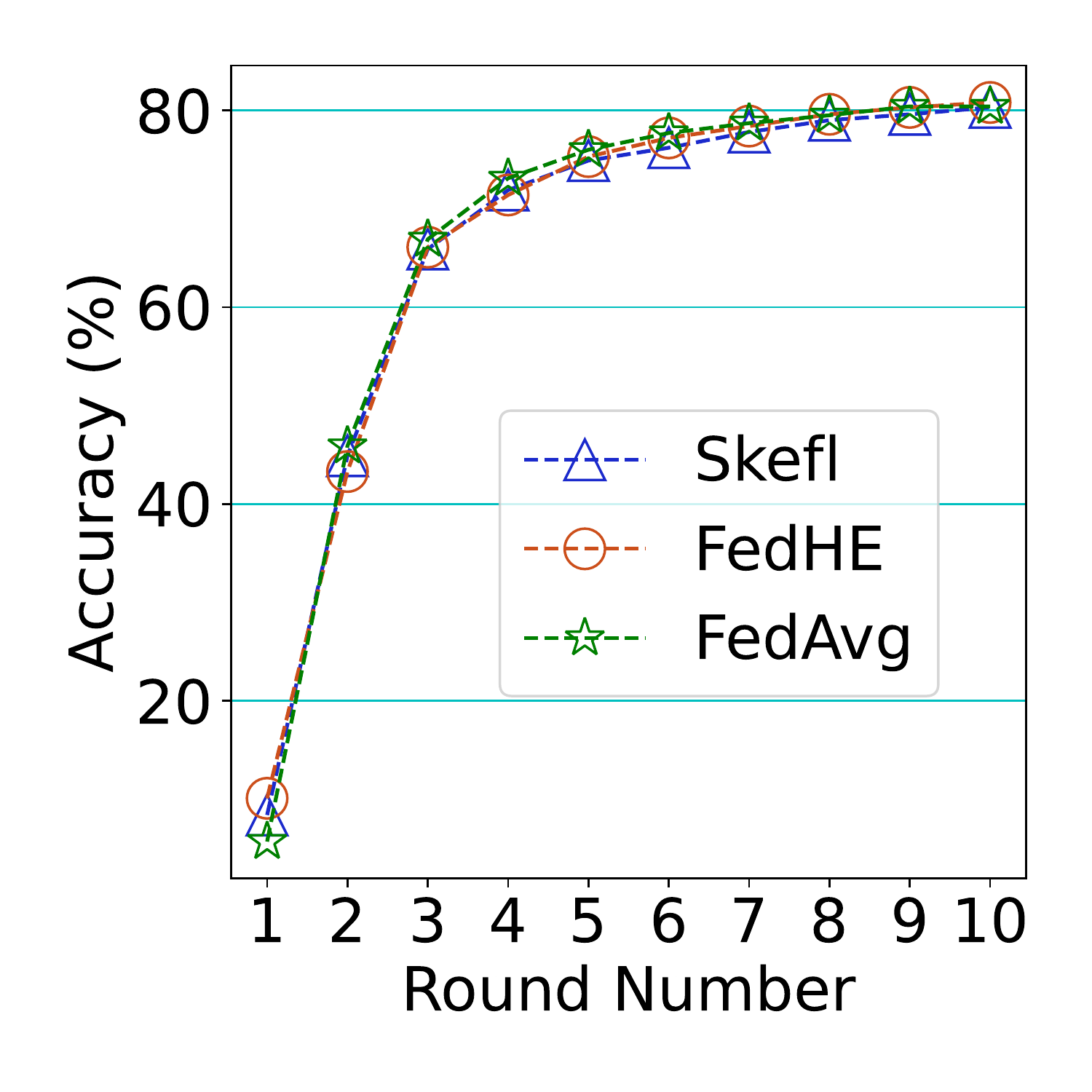}
     \end{subfigure}
     \hfill
     \begin{subfigure}[b]{\subfigwidth\textwidth}
         \centering
         \includegraphics[width=\textwidth]{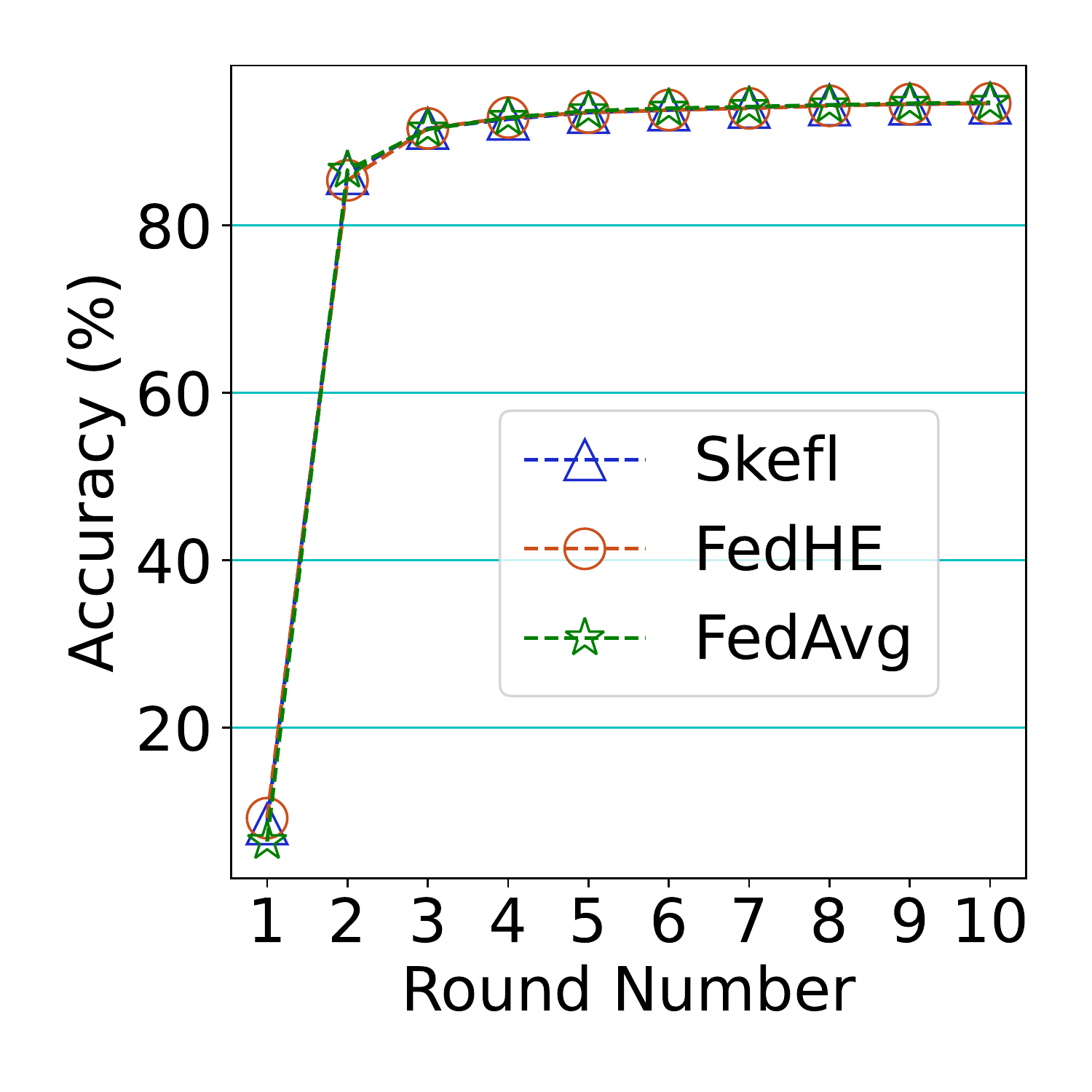}
     \end{subfigure}     
        \caption{Skefl accuracy for MNIST, FMNIST, CIFAR-10, and SVHN}
        \label{fig:accuracy}
\end{figure*}

\begin{figure*}[!t]
     \centering
     \begin{subfigure}[b]{\subfigwidth\textwidth}
         \centering
         \includegraphics[width=\textwidth]{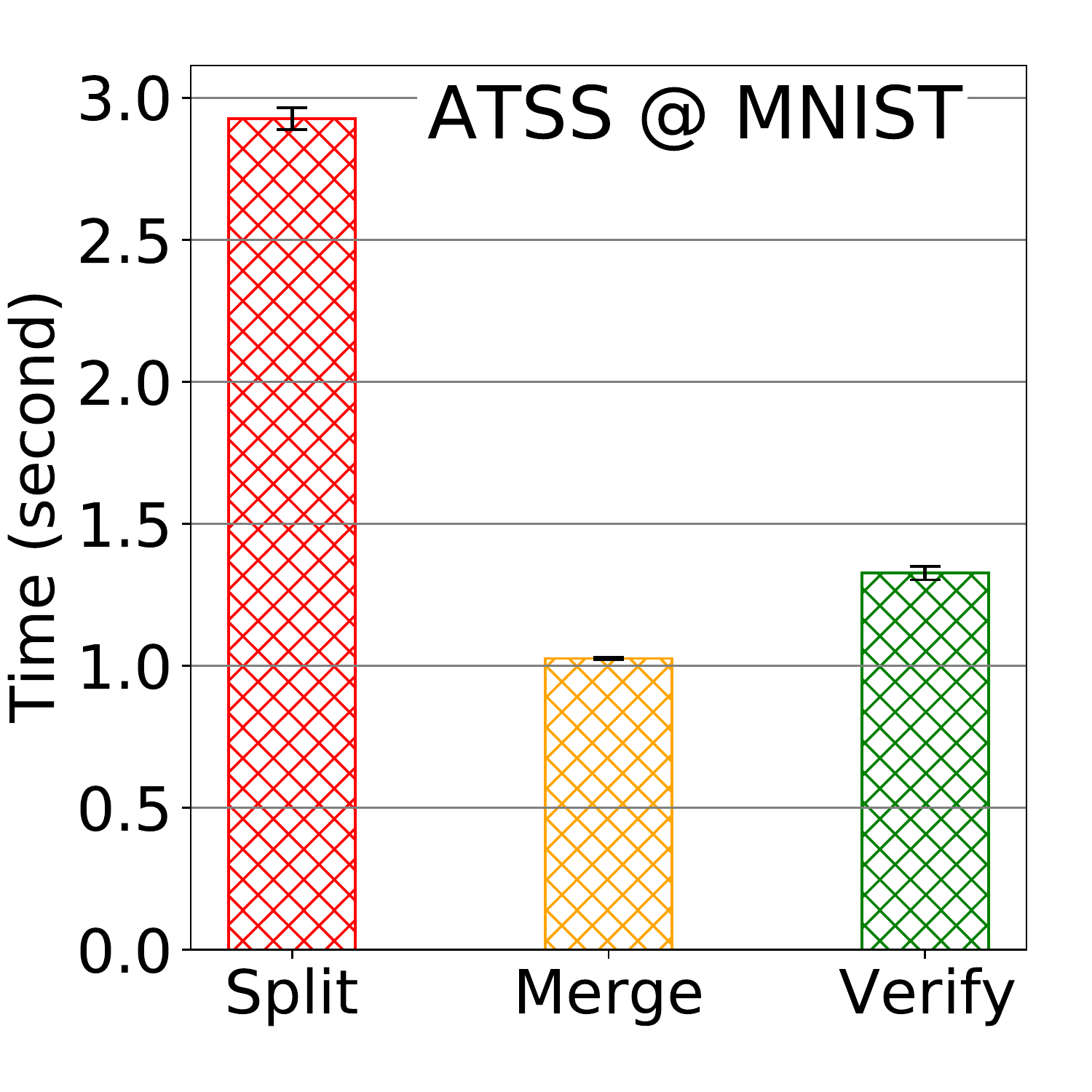}
     \end{subfigure}
     \hfill
     \begin{subfigure}[b]{\subfigwidth\textwidth}
         \centering
         \includegraphics[width=\textwidth]{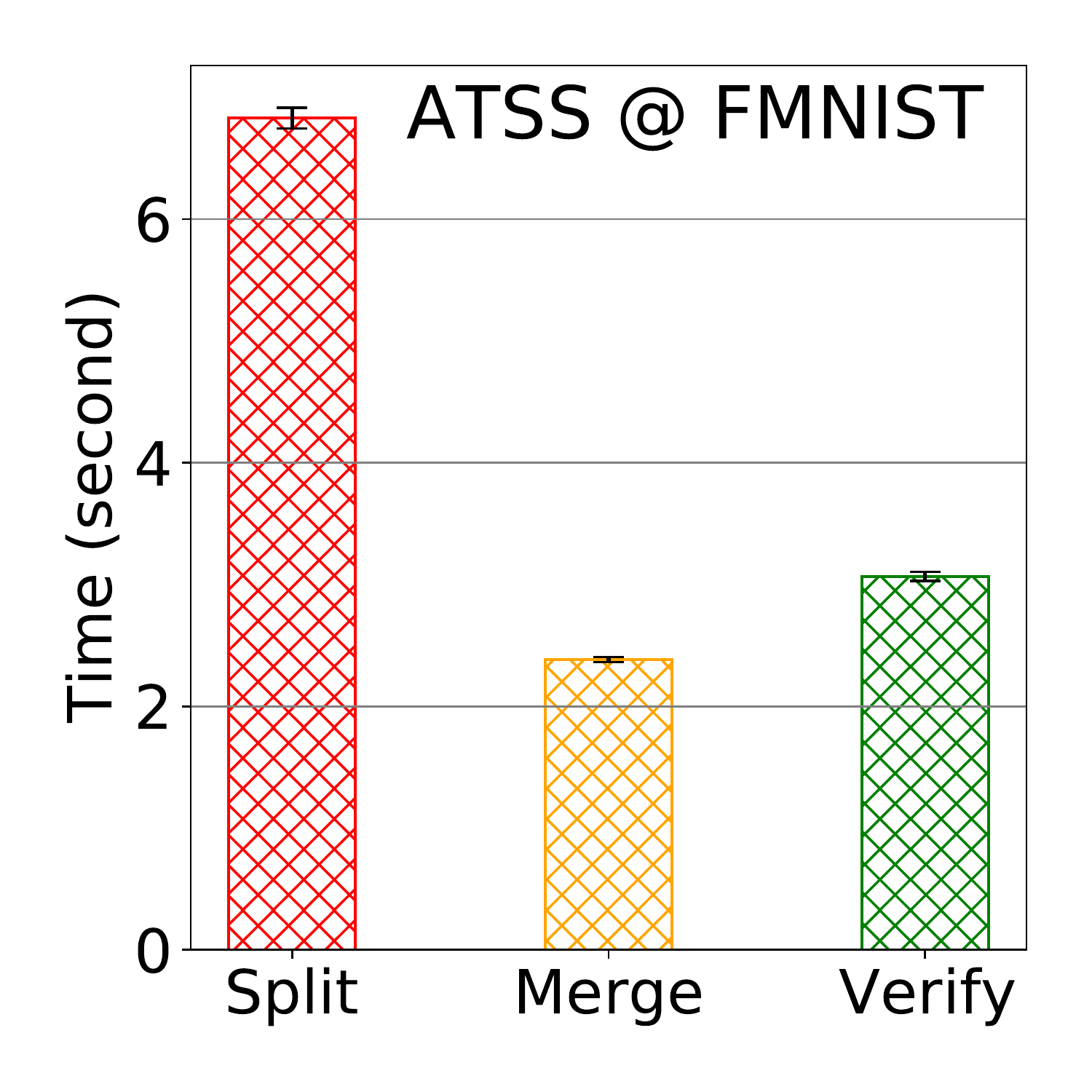}
     \end{subfigure}
     \hfill
     \begin{subfigure}[b]{\subfigwidth\textwidth}
         \centering
         \includegraphics[width=\textwidth]{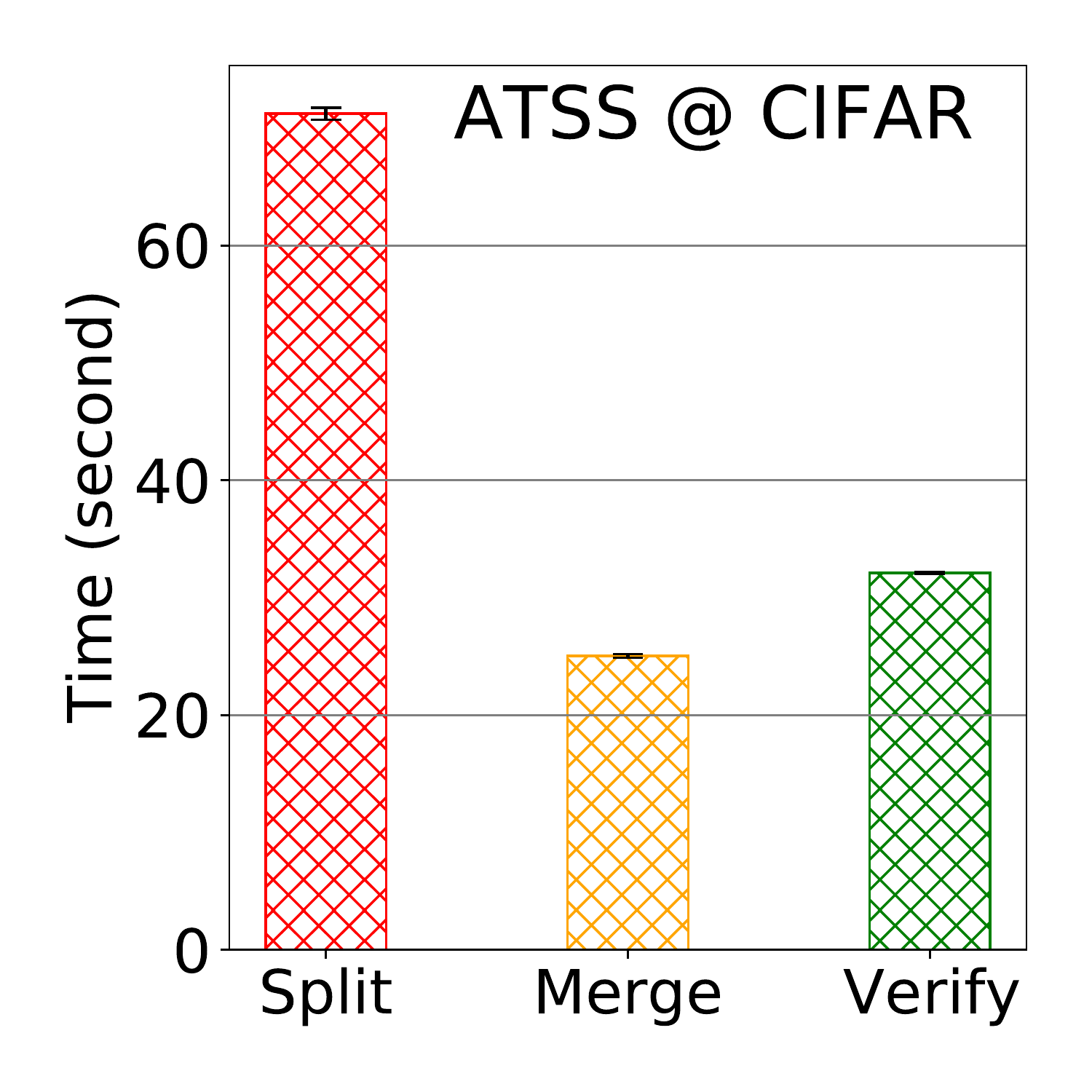}
     \end{subfigure}
     \hfill
     \begin{subfigure}[b]{\subfigwidth\textwidth}
         \centering
         \includegraphics[width=\textwidth]{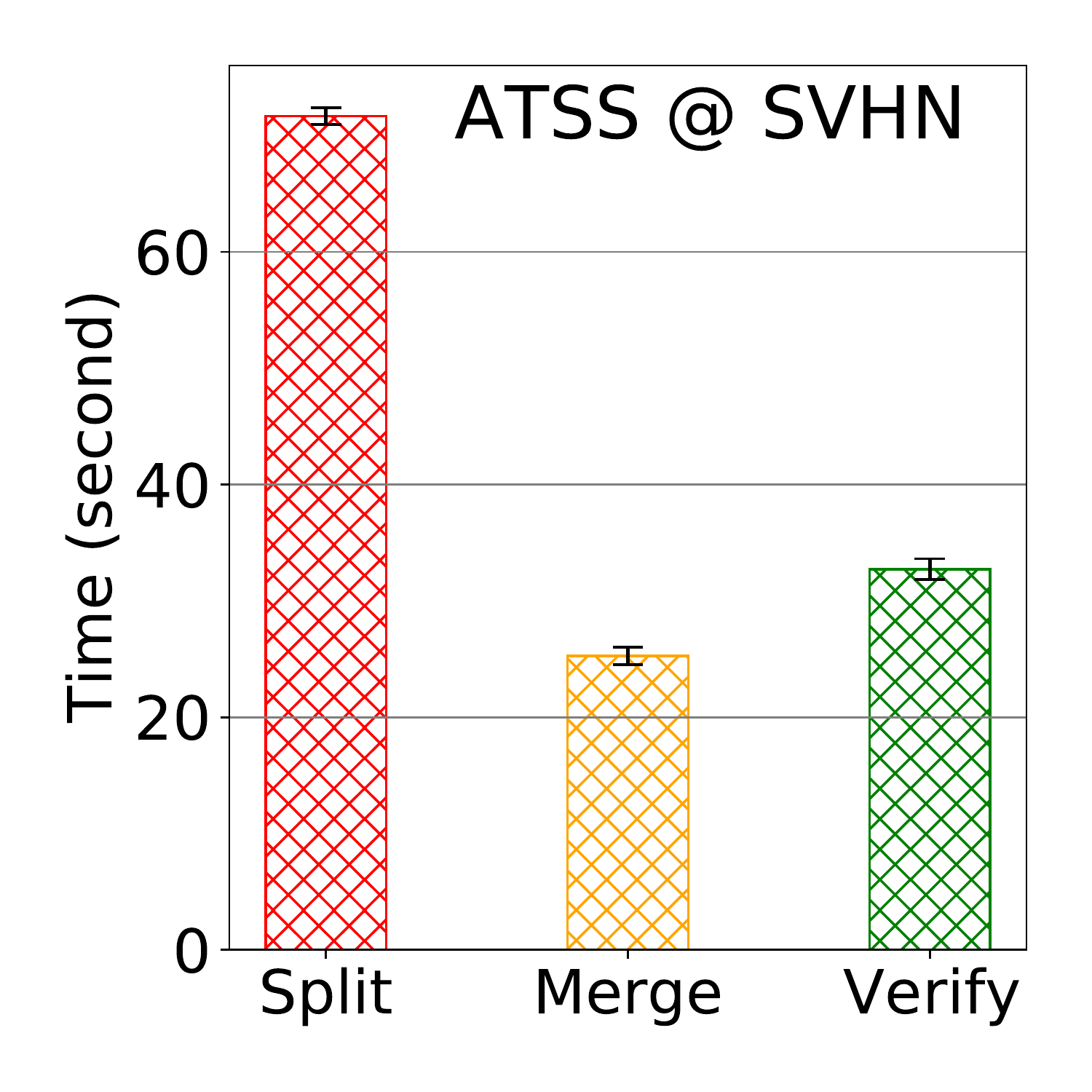}
     \end{subfigure}     
        \caption{Performance of ATSS primitives on MNIST, FMNIST, CIFAR-10, and SVHN}
        \label{fig:atts}
\end{figure*}

\begin{figure*}[!t]
     \centering
     \begin{subfigure}[b]{\subfigwidth\textwidth}
         \centering
         \includegraphics[width=\textwidth]{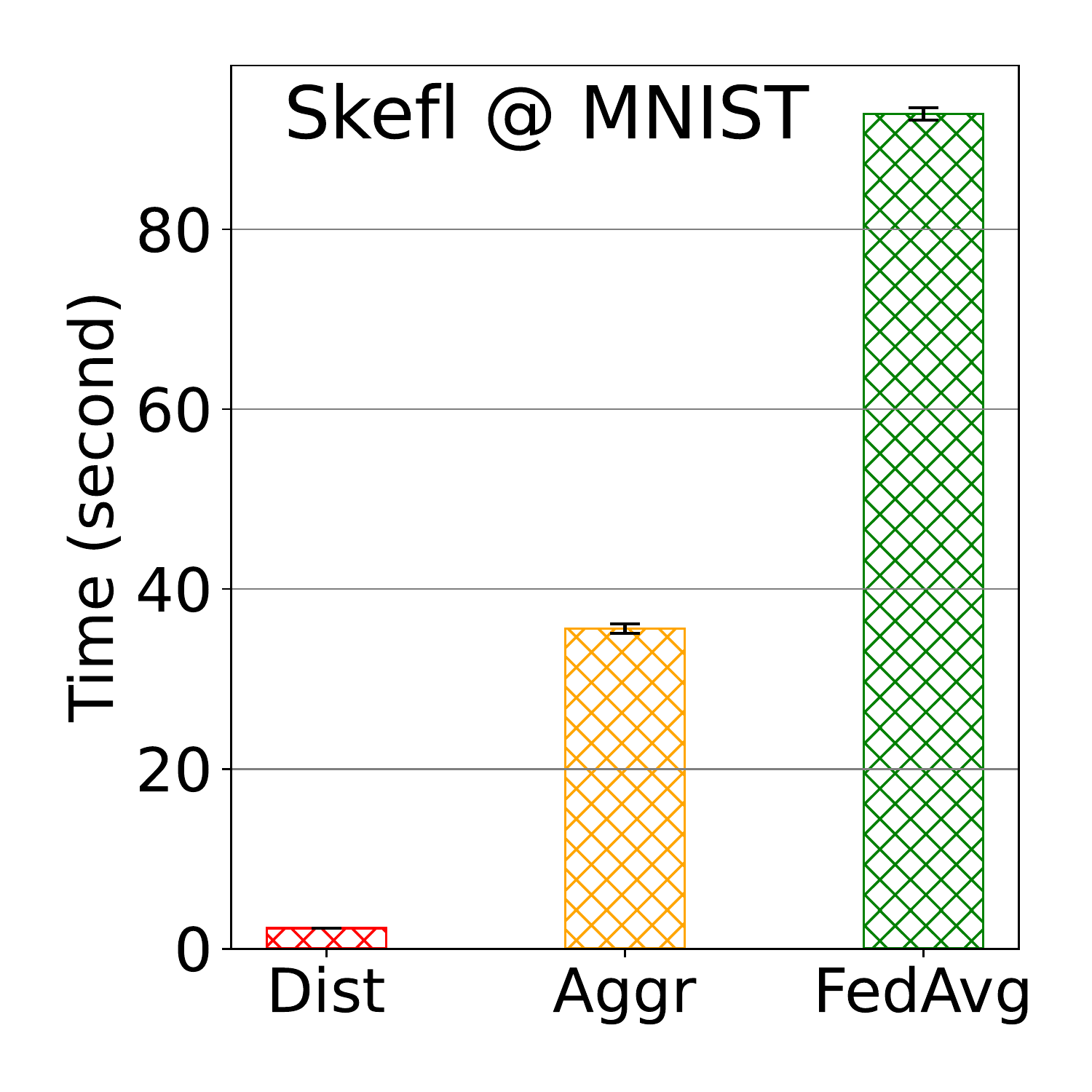}
     \end{subfigure}
     \hfill
     \begin{subfigure}[b]{\subfigwidth\textwidth}
         \centering
         \includegraphics[width=\textwidth]{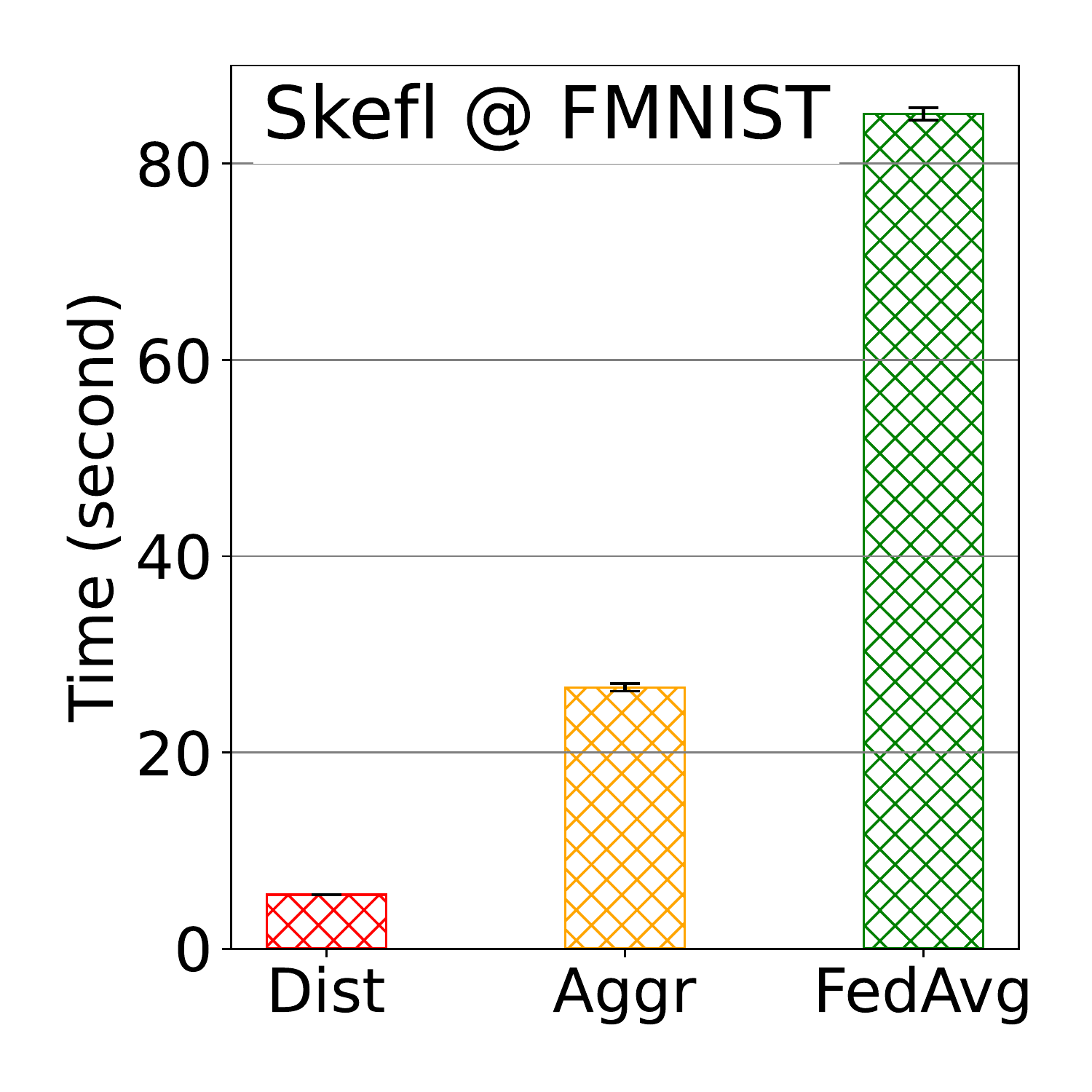}
     \end{subfigure}
     \hfill
     \begin{subfigure}[b]{\subfigwidth\textwidth}
         \centering
         \includegraphics[width=\textwidth]{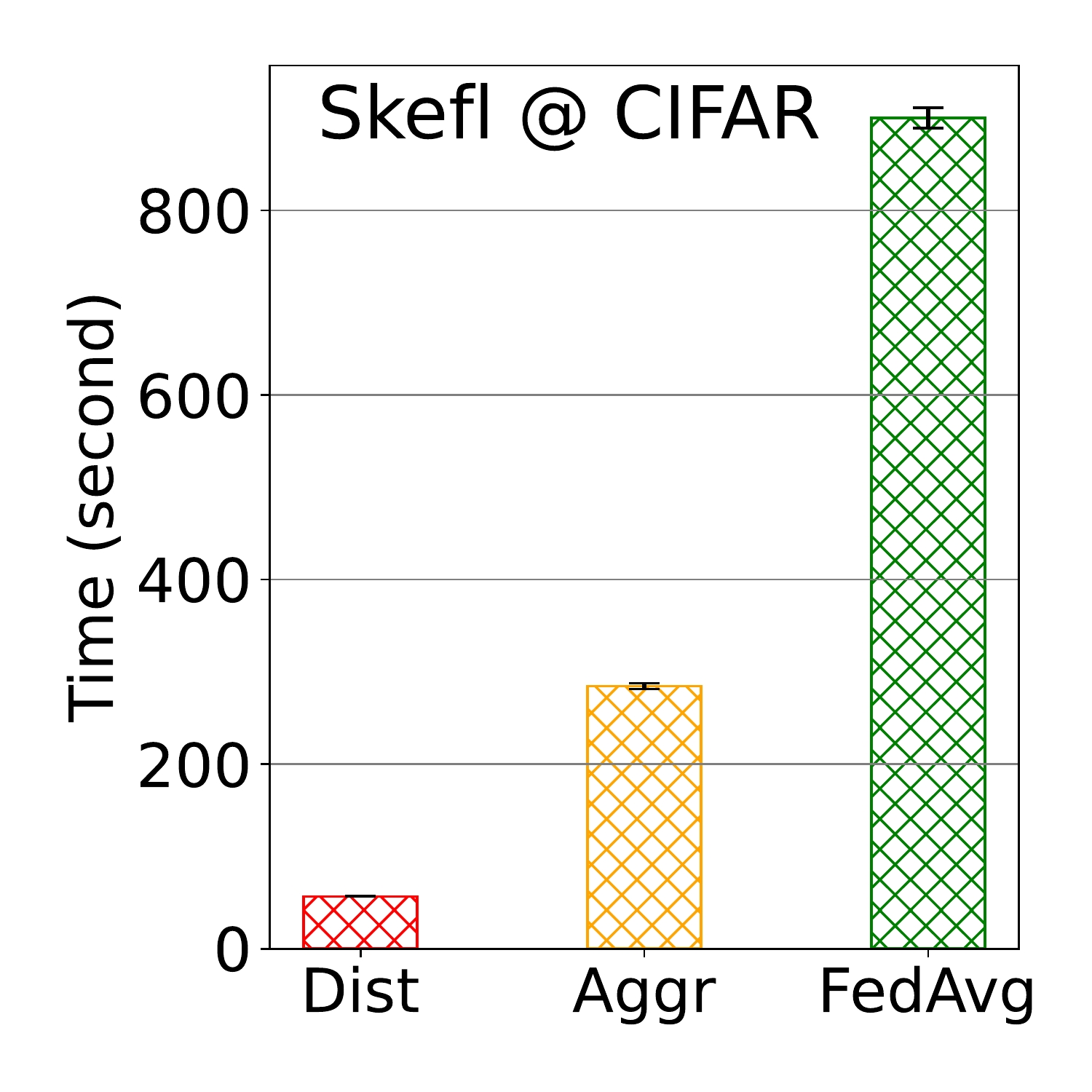}
     \end{subfigure}
     \hfill
     \begin{subfigure}[b]{\subfigwidth\textwidth}
         \centering
         \includegraphics[width=\textwidth]{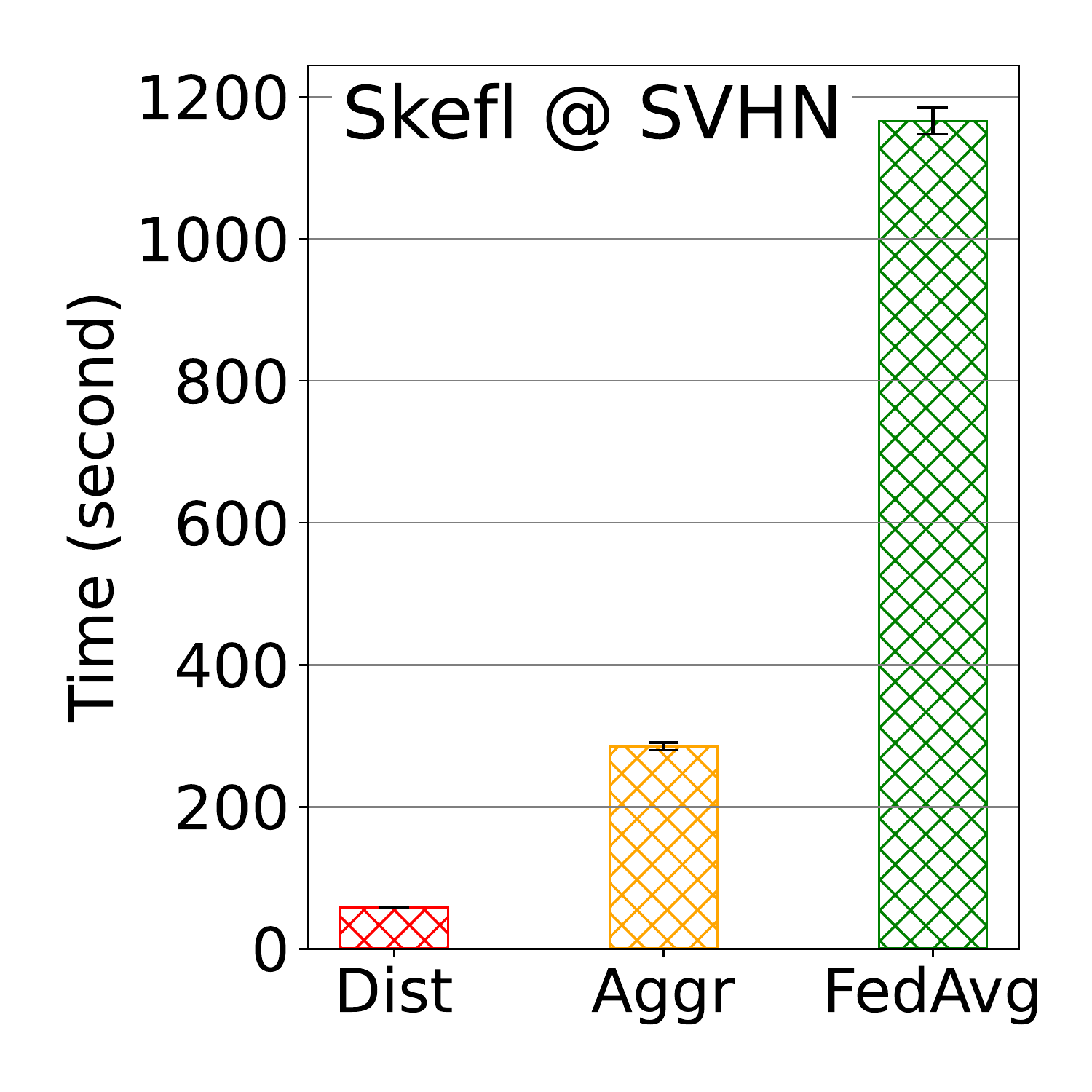}
     \end{subfigure}     
        \caption{Skefl overhead of FedAvg on MNIST, FMNIST, CIFAR-10, and SVHN}
        \label{fig:fedsh}
\end{figure*}

\subsubsection{Overall Accuracy}

In this subsection, we delve into the heart of our experimental evaluation by presenting the overall accuracy performance of the proposed Skefl protocol within the context of the Federated Machine Learning (FedML) paradigm. We engage in a rigorous comparison, pitting the accuracy performance of Skefl against two well-established baseline aggregation algorithms: FedAvg and FedHE. The culmination of these investigations is vividly depicted in Fig.~\ref{fig:accuracy}, where the trajectory of accuracy is meticulously traced over multiple rounds of aggregation. The graphical representation unequivocally showcases the convergence trends, revealing that Skefl, FedAvg, and FedHE all exhibit rapid convergence, achieving accuracy levels that soar past the 80\% threshold within a remarkably scant 10 aggregation rounds. This compelling outcome unambiguously underscores the potency of Skefl in adeptly harmonizing the twin objectives of accuracy preservation and privacy assurance during the aggregation process. The capability of Skefl to uphold accuracy at a competitive level vis-\`a-vis conventional methodologies stands as a compelling testament to its pragmatic utility and relevance.

\subsubsection{ATSS Performance}

Transitioning to the meticulous assessment of the Asymmetric Threshold Secret Sharing (ATSS) primitives, meticulously delineated in Fig.~\ref{fig:atts}, we meticulously scrutinize the temporal dynamics of the execution times associated with the three constituent ATSS primitives: \textbf{ATSS.Split}(), \textbf{ATSS.Combine}(), and \textbf{ATSS.Aggregate}(). The observed trends in execution times delineate a discernible pattern. For datasets characterized by relative simplicity, exemplified by MNIST and FMNIST, all three ATSS primitives promptly culminate their execution in a matter of mere seconds. Nevertheless, the narrative changes when grappling with more intricate datasets, typified by CIFAR-10 and SVHN, where execution times elongate to span tens of seconds. Evidently, \textbf{ATSS.Split}() emerges as the harbinger of the longest execution durations among the ATSS primitives. 

\subsubsection{Skefl Overhead}

Embarking on a distinct avenue of analysis, we delve into the nuanced exploration of the overhead incurred by the Skefl protocol, encapsulated within Fig.~\ref{fig:fedsh}. This analysis delves into the computational costs borne by the two pivotal functions embedded within the Skefl protocol: \textbf{Skefl\_Dist}() and \textbf{Skefl\_Aggr}(). The findings elegantly elucidate the delicate equilibrium that characterizes the trade-off between the gains in privacy bestowed by secret sharing and the attendant computational overhead. Astonishingly, \textbf{Skefl\_Dist}() shines forth as a paragon of efficiency, incurring negligible overhead that essentially aligns with the streamlined efficiency of the pristine FedAvg protocol. On a divergent trajectory, \textbf{Skefl\_Aggr}() exacts a relatively heftier computational toll, accounting for a range spanning from 24\% to 38\% of the temporal expenditure entailed by the analogous FedAvg operation.

\section{Conclusion}

Conventional single-key HE schemes used in federated learning assume non-collusion between the parameter server and participating clients, a vulnerability wherein an adversary client could compromise the local model of another by intercepting ciphertexts. To address this, our proposed Skefl protocol extends single-key HE schemes with efficient secret sharing, ensuring that collusion between the parameter server and any compromised clients does not reveal local models. The security of Skefl is rigorously proven using a well-established simulation framework in cryptography. Additionally, the practical performance of Skefl is reported, highlighting its effectiveness in achieving a robust balance between security and efficiency in secure federated learning scenarios.

\bibliographystyle{named}
\bibliography{ijcai24}

\end{document}